\documentclass[twocolumn,aps,prb,superscriptaddress]{revtex4-2}
\usepackage{graphicx}
\usepackage{amsmath}

\begin{document}

\title{Topological Electronic Structure Evolution with Symmetry Breaking Spin Reorientation in (Fe$_{1-x}$Co$_{x}$)Sn }

\author{Robert G. Moore}
\email{moorerg@ornl.gov\newline
This manuscript has been authored in part by UT-Battelle, LLC, under contract DE-AC05-00OR22725 with the US Department of Energy (DOE). The US government retains and the publisher, by accepting the article for publication, acknowledges that the US government retains a nonexclusive, paid-up, irrevocable, worldwide license to publish or reproduce the published form of this manuscript, or allow others to do so, for US government purposes. DOE will provide public access to these results of federally sponsored research in accordance with the DOE Public Access Plan (http://energy.gov/downloads/doe-public-access-plan)}
\affiliation{Materials Science and Technology Division, Oak Ridge National Laboratory, Oak Ridge, TN 37831, USA}
\author{Satoshi Okamoto}
\affiliation{Materials Science and Technology Division, Oak Ridge National Laboratory, Oak Ridge, TN 37831, USA}
\author{Haoxiang Li}
\affiliation{Materials Science and Technology Division, Oak Ridge National Laboratory, Oak Ridge, TN 37831, USA}
\author{William R. Meier}
\affiliation{Materials Science and Technology Division, Oak Ridge National Laboratory, Oak Ridge, TN 37831, USA}
\author{Hu Miao}
\affiliation{Materials Science and Technology Division, Oak Ridge National Laboratory, Oak Ridge, TN 37831, USA}
\author{Ho Nyung Lee}
\affiliation{Materials Science and Technology Division, Oak Ridge National Laboratory, Oak Ridge, TN 37831, USA}
\author{Makoto Hashimoto}
\affiliation{Stanford Synchrotron Radiation Lightsource, SLAC National Accelerator Laboratory, Menlo Park, CA 94025, USA}
\author{Donghui Lu}
\affiliation{Stanford Synchrotron Radiation Lightsource, SLAC National Accelerator Laboratory, Menlo Park, CA 94025, USA}
\author{Elbio Dagotto}
\affiliation{Materials Science and Technology Division, Oak Ridge National Laboratory, Oak Ridge, TN 37831, USA}
\affiliation{Department of Physics and Astronomy, University of Tennessee, Knoxville, TN 37996, USA}
\author{Michael A. McGuire}
\affiliation{Materials Science and Technology Division, Oak Ridge National Laboratory, Oak Ridge, TN 37831, USA}
\author{Brian C. Sales}
\affiliation{Materials Science and Technology Division, Oak Ridge National Laboratory, Oak Ridge, TN 37831, USA}

\begin{abstract}

Topological materials hosting kagome lattices have drawn considerable attention due to the interplay between topology, magnetism, and electronic correlations.  The (Fe$_{1-x}$Co$_x$)Sn system not only hosts a kagome lattice but has a tunable symmetry breaking magnetic moment with temperature and doping.  In this study, angle resolved photoemission spectroscopy and first principles calculations are used to investigate the interplay between the topological electronic structure and varying magnetic moment from the planar to axial antiferromagnetic phases.  A theoretically predicted gap at the Dirac point is revealed in the low temperature axial phase but no gap opening is observed across a temperature dependent magnetic phase transition.  However, topological surface bands are observed to shift in energy as the surface magnetic moment is reduced or becomes disordered over time during experimental measurements.  The shifting surface bands may preclude the determination of a temperature dependent bulk gap but highlights the intricate connections between magentism and topology with a surface/bulk dichotomy that can affect material properties and their interrogation.

\end{abstract}

\maketitle

\date{\today }

\section{INTRODUCTION}

Topological insulators and semimetals have received significant attention due to their unique linearly dispersive massless Dirac electronic bands that are topologically protected~\cite{MooreJ2010, Hasan2010, Yan2017, ARmitage2018}.  The breaking of symmetry by introducing magnetism into topological systems expands the exotic states that can be created including the quantum anomalous Hall state, axionic insulating state, and chiral anomalies to name a few~\cite{Chang2013, Liu2020, Mong2010, Hirschberger2016}.  Breaking symmetry by the creation of a surface or interface also controls the interplay between magnetism and topology as, for example, the two-dimensional topological surface states observed in the AFM phases of GdSbTe~\cite{Hosen2018}, EuSn$_2$As$_2$~\cite{Li2019} and MnBi$_2$Te$_4$~\cite{Li2019, Otrokov2019, Chen2019, Hao2019, Vidal2019, Lee2019}. 

While theory predicts specific implications after the breaking of time reversal symmetry on the topologically protected electronic states, such as the creation of a gap at the Dirac point, experimental observations are less clear.  For example, there are conflicting reports on the existence~\cite{Otrokov2019, Vidal2019, Lee2019} or absence~\cite{Li2019, Chen2019, Hao2019} of an exchange gap in the topological states of MnBi$_2$Te$_4$ as well as the coexistence of surface magnetism without a gap~\cite{Nevola2020} and the creation of a gap only through tuning the Fermi level~\cite{Ko2020}.    Hence, finding systems where the magnetism can be tuned is important for understanding and controlling emergent behavior in topological systems and delineating the different effects in the bulk and on the surface.

Topological materials with kagome lattices are of particular interest due to the emergence of dispersionless flat electronic bands and saddle points in addition to linearly dispersive bands~\cite{Lin2018, Ye2018, Liu2018, Kuroda2017,Meier2020}.  In addition, magnetic kagome materials can host both ferromagnetic (FM) and antiferromagnetic (AFM) spin arrangements.  For example, both Fe$_3$Sn$_2$~\cite{Ye2018} and Co$_3$Sn$_2$S$_2$~\cite{Liu2018} contain FM kagome lattices while Mn$_3$Sn~\cite{Kuroda2017} displays an AFM ordering.  The magnetic topological materials with kagome lattices adds magnetic frustration and large density of electronic states to systems with topological protection.  It has been shown that the spin arrangement on the magnetic kagome lattice of the (Fe$_{1-x}$Co$_x$)Sn system can be controlled with both composition and temperature~\cite{Meier2019}.  Thus the (Fe$_{1-x}$Co$_x$)Sn system provides an ideal opportunity to tune the magnetic state while observing the interplay between magnetism and topologically protected electronic states.

FeSn forms in a hexagonal structure (space group P6/mmm, No. 191) consisting of kagome Fe$_3$Sn layers separated by Sn sheets schematically shown in Fig. 1a~\cite{Meier2019, Djega1969, Haggstrom1975}.    FeSn has itinerant FM magnetic moments within the kagome lattice and parallel to the kagome plane which are aligned antiparallel with neighboring layers along the c-axis to form a bulk AFM arrangement~\cite{Meier2019, Haggstrom1975, Yamaguchi1967, Kulshreshtha1981}.  While the magnetic arrangement breaks inversion (P) and time reversal (T) symmetry, the combination of inversion plus time-reversal (PT) symmetry is preserved along with a nonsymmorphic magnetic twofold screw rotation symmetry ($S_{2z}$) along the c-axis~\cite{Lin2020}.  Band structure calculations including spin-orbit coupling show that the combined PT symmetry protects massless Dirac points appearing around $E-E_F\sim 0.4$ eV at the H and H’ points.  These Dirac points (DP) can be viewed as a pair of degenerate Weyl points (WP) due to AFM coupling of neighboring FM kagome lattices.  On the surface of FeSn, a Stark effect occurs for the topmost Fe$_3$Sn layer and lifts the degeneracy of the Weyl points and shifts the surface kagome layer Weyl band to $E-E_F\sim0.25$ eV in energy.    The breaking of the combined PT symmetry on the surface results in a massive surface Weyl band with a predicted small $\sim5$ meV gap~\cite{Lin2020, You2019, Wu2019}.  

\begin{figure}
\begin{center}
\includegraphics[width=1\columnwidth, clip]{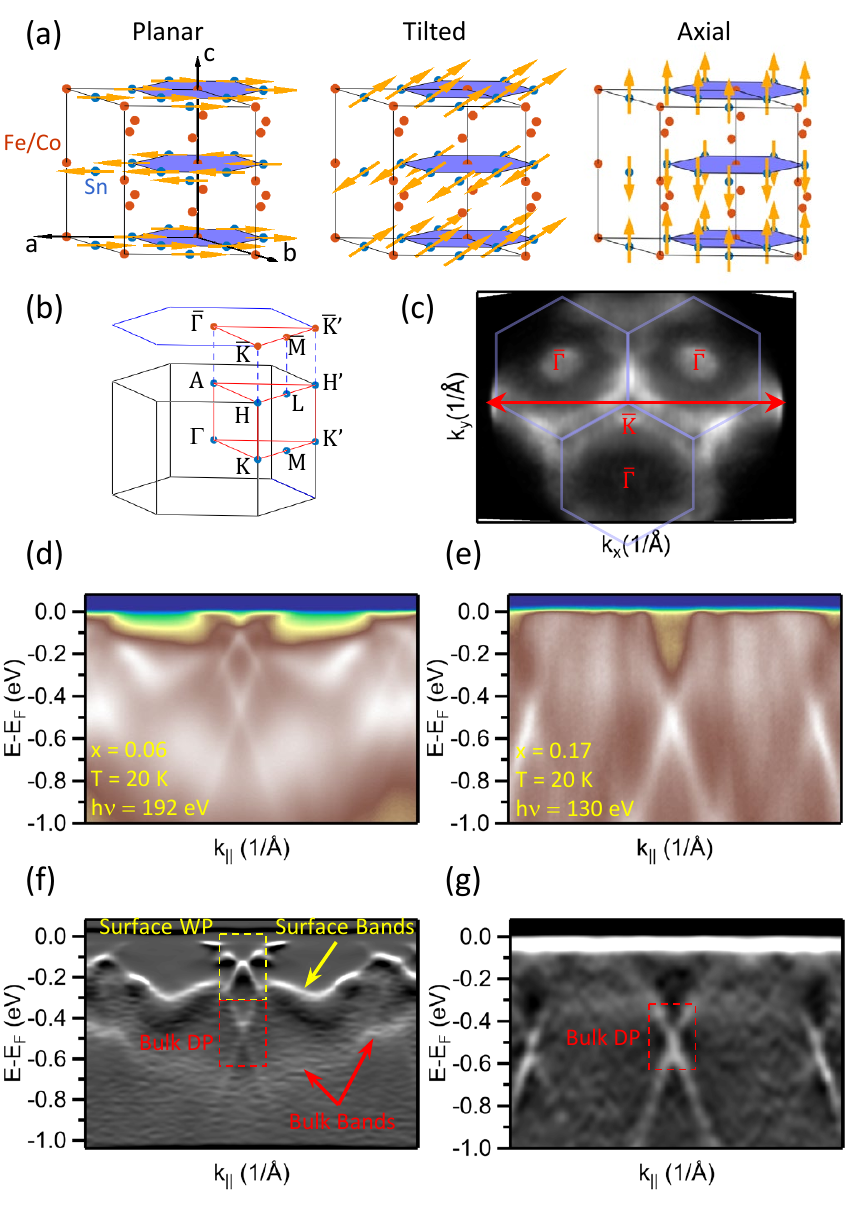}
\caption{Magnetic and electronic structure of (Fe$_{1-x}$Co$_x$)Sn.  (a) Schematic of atomic and magnetic structure for the planar, tilted and axial magnetic phases.  The kagome lattice is highlighted in purple. (b) Brillouin zone (BZ) with surface projection and high symmetry points identified.  (c) Fermi surface map for $x = 0.06$ with (BZ) overlay.  Red arrow indicates ARPES energy-momentum cut for subsequent data. (d) ARPES data for $x = 0.06$ at $h\nu = 192$ eV, chosen to maximize the Dirac and Weyl band intensities.  (e) ARPES data for $x = 0.17$ at $h\nu = 130$ eV which corresponds to the H and H’ points where a maximum gap at the Dirac point is predicted.  (f), (g) Curvature plot for ARPES data in (d) and (e) respectively.  Surface WP for $x = 0.06$ is outlined in yellow and bulk DPs are outlined in red.}
\label{Overview}
\end{center}
\end{figure}

The addition of Co into the FeSn system to create (Fe$_{1-x}$Co$_x$)Sn results in a reorientation of the itinerant magnetic moments in the kagome layers~\cite{Meier2019}.  While the moments within each layer maintain their FM alignment, increasing Co tilts the magnetic moments from a planar phase parallel to the kagome plane through a tilted phase with magnetic moments tilted out of the kagome plane to an axial phase with magnetic moments oriented along the c-axis.  Despite the reorientation of the magnetic moments in the kagome planes, neighboring layers maintain their AFM arrangement as shown schematically in Fig. 1a.  For $x\sim0.15$ the axial phase is the ground state of the system but for $0 < x < 0.15$, a temperature dependent second order phase transition between the phases is observed~\cite{Meier2019}.  Theoretical calculations have predicted that this spin reorientation to the axial phase should break the nonsymmorphic $S_{2z}$ symmetry and open a large $\sim 70$ meV gap at the H and H’ Dirac points in the electronic dispersion~\cite{Lin2020}.

Here angle resolved photoemission spectroscopy (ARPES) combined with constrained magnetic moment first principles calculations is used to investigate the evolution of the electronic structure of (Fe$_{1-x}$Co$_x$)Sn with both composition and temperature.  The bulk Dirac and surface Weyl bands are clearly observed with a gap at the bulk Dirac point in the low temperature axial ground state, in agreement with theoretical calculations.  However, no changes are observed in the experimental bulk Dirac dispersion as the magnetic phase is tuned with temperature from planar through tilted to the axial magnetic phase, which disagrees with theoretical predictions.   A distinct evolution of the surface Weyl bands compared to the bulk Dirac bands is observed indicative of a reduction or disordering of the magnetic moment on the surface, which is likely due to sample aging of the cleaved bulk crystals.  The shifting of the Weyl band energies combined with thermal broadening of the data could obscure the Dirac gap evolution with temperature but the interplay between magnetism and topology is evident.  Further experimental and theoretical investigation of the surface magnetism for the different surface terminations is required to fully understand the origins of the observed surface band evolution. Nonetheless, the dichotomy between surface and bulk behavior highlights the challenges in interrogating magnetic topological systems that can lead to contradicting results.

\section{METHODS}

Crystals of (Fe$_{1-x}$Co$_x$)Sn were grown via the flux method, i.e. slow cooling a melt with a tin flux.  The crystal structure, transport and magnetic properties were well characterized prior to the photoemission measurements as described elsewhere~\cite{Meier2019}.  For these studies, $x = 0.06$ and $0.17$ were used to investigate the temperature dependent phases and axial ground state, respectively.  

The ARPES measurements were conducted at Beamline 5-2 at the Stanford Synchrotron Radiation Lightsource utilizing a Scienta DA30L electron spectrometer and base pressure of better than $5\times10^{-11}$ Torr.  Linearly horizontal polarized light in the photon energy range $h\nu = 90-130$ eV were used for the measurements.  The beamline has a nominal $0.04$ mm$^2$ spots size and a total energy resolution $\sim16$ meV was set for the measurements.  Single crystals were mounted to sample posts using silver epoxy and cleaved in vacuum by knocking off a post mounted to the top of the sample.

To examine the dependence of the electronic band structure on magnetic ordering, we carry out Density Functional Theory (DFT) calculations. Following the DFT work by Sales et al., ~\cite{Sales2019} we use the Vienna {\it ab initio} simulation package (VASP) ~\cite{Kresse1996}, which uses the projector augmented wave method ~\cite{Kresse1999} with the generalized gradient approximation in the parametrization of Perdew, Burke, and Enzerhof ~\cite{Perdew1996} for the exchange-correlation. For Fe a standard potential is used (Fe in the VASP distribution), and for Sn a potential where the {\it d} states are treated as valence states, is used (Sn$_d$). We use the experimental structure doubled along the c-axis direction to account for the layered AFM ordering. In most cases, we use an $8 \times 8 \times 6$ k-point grid and an energy cutoff of $500$ eV. The $+U$ correction is not included because FeSn is an itinerant magnetic system, but the spin-orbit coupling (SOC) is included to study the dependence of the electronic band structure on the direction as well as the size of ordered moments on Fe sites. For this purpose, we carry out the constrained magnetism calculations by setting \texttt{I\char`_CONSTRAINED\char`_M} $= 2$ and fixing the direction and the size of ordered moments.

\section{RESULTS}

The ARPES Fermi surface for (Fe$_{1-x}$Co$_x$)Sn $x = 0.06$ is shown in Fig.~\ref{Overview}c and is similar to previous reports ~\cite{Lin2020}.  Figure~\ref{Overview}d and ~\ref{Overview}e show the low temperature electronic structure for $x = 0.06$ and $0.17$ respectively, through the $\bar{\rm K}$ point in a direction that is perpendicular to the $\Gamma  - K$ direction.  This cut orientation was chosen to maximize the appearance of the bulk Dirac and surface Weyl bands.  The topological bands are clearly resolved in the raw data but Fig.~\ref{Overview}f and ~\ref{Overview}g show plots of a ``curvature" analysis highlighting the dispersive features~\cite{Zhang2011}.  The (Fe$_{1-x}$Co$_x$)Sn system has a three dimensional electronic dispersion and different $k_z$ are accessible by varying the photon energy.  From previous reports, $h\nu = 130$ eV corresponds to the $k_z = \pi/c$, where the H and H’ points occur~\cite{Lin2020}.  Figure ~\ref{Overview}e shows the electronic structure at $h\nu = 130$ eV for $x = 0.17$ in the axial ground state.  While the bulk Dirac bands are clearly visible in the data, photon energy dependent matrix elements suppress the observation of the surface Weyl bands.  However, using $h\nu = 92$ eV, which corresponds to $k_z \sim 0.48 \pi/c $, both the bulk Dirac bands and surface Weyl bands are clearly visible as shown for $x = 0.06$ in Fig.~\ref{Overview}d.  For $x = 0.06$ the bulk Dirac point (DP) and surface Weyl point (WP) are located at $E-E_F\sim-0.41$ eV and $E-E_F \sim -0.12$ eV respectively in agreement with previous measurements of undoped FeSn~\cite{Lin2020}.  For $x = 0.17$ the WP intensity suppression prevents an accurate determination, but the DP is located further below the Fermi energy at $E-E_F \sim -0.5$ eV. 

\begin{figure*}[t]
\begin{center}
\includegraphics[keepaspectratio=true, width = 7.0 in]{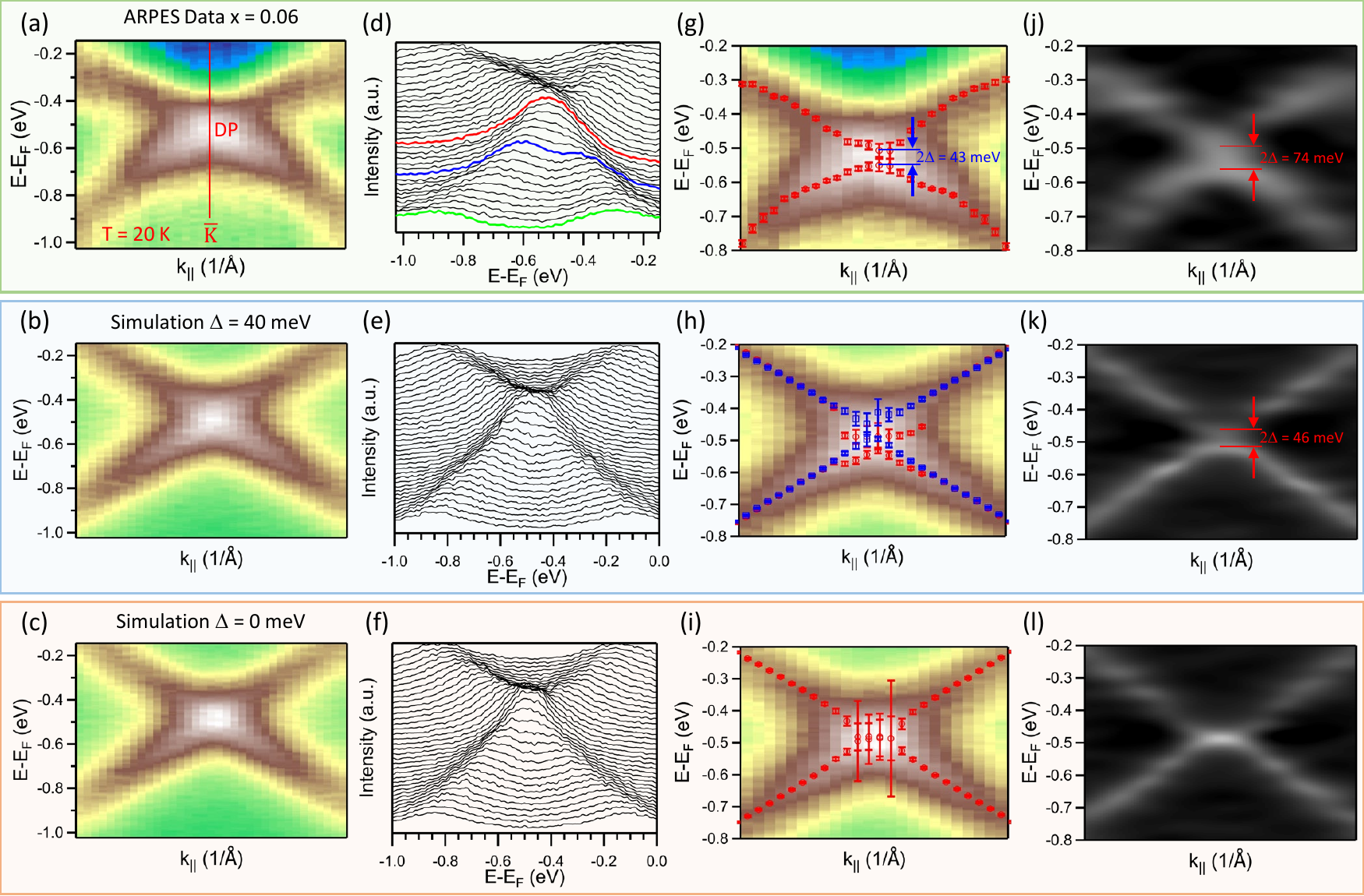}
\caption{Comparison of ARPES data for $x = 0.17$ and the low temperature axial phase with simulations. (a-c) Image plots for region around the bulk DP outlined in Fig.~\ref{Overview}g for binned ARPES data, $\Delta = 40$ meV simulation and $\Delta = 0$ meV simulation respectively.  (d-f) EDC curves for binned ARPES data, $\Delta = 40$ meV and $\Delta = 0$ meV simulations respectively.  For the ARPES EDCs the red, blue and green curves highlight regions with different intensities and widths (see main text).  (g-i) Fitted dispersion results for ARPES data, $\Delta = 40$ meV, and $\Delta = 0$ meV simulations respectively.  The blue markers are for free fit parameters while the red markers are for constrained fits.  (j-l) Curvature analysis for ARPES data, $\Delta = 40$ meV and $\Delta = 0$ meV simulations respectively.}
\label{Simulations}
\end{center}
\end{figure*}

\subsection{Axial Ground State for $x = 0.17$}
Theoretical band structure calculations predict the axial phase to open a large $2\Delta \sim70$ meV gap at the bulk Dirac point in the axial phase~\cite{Lin2020}.  To understand the implications of the axial AFM phase on the electronic structure, a closer investigation of the Dirac bands highlighted in Fig.~\ref{Overview}g are presented in Fig.~\ref{Simulations}.  The curvature analysis for the data in Fig.~\ref{Overview}g around the vicinity of the Dirac point, displayed in Fig.~\ref{Simulations}j, shows a gapped electronic structure with $2\Delta \sim 74$ meV.  However, since such features are hard to observe in the raw data, an in-depth investigation is performed to confirm the reliability of such an analysis. 

Figure~\ref{Simulations}a shows the data in the vicinity of the Dirac point that has been binned along the momentum axis to aid in the analysis.  Plots of the energy spectra at constant momentum, i.e. Energy Distribution Curves (EDCs), are presented for each bin in Fig.~\ref{Simulations}d.  Several features in the data and EDCs should be noted that can affect the analysis and our understanding of the existence, or not, of a gap at the Dirac point.  

From the data plotted in Fig.~\ref{Overview}e as well as the plot of EDCs in Fig.~\ref{Simulations}d, the intensity of the Dirac bands above the Dirac point is markedly lower compared to the bands below the Dirac point.  Upon closer inspection, the intensity of the bands above and below the Dirac point are similar away from the Dirac point, as highlighted by the green curve in Fig.~\ref{Simulations}d and Fig. S2 in the Supplemental Material~\cite{SM}, and only the intensity of the bands above the Dirac point are suppressed as the Dirac point is approached, emphasized by the blue curve in Fig.~\ref{Simulations}d (see also Fig. S2~\cite{SM}).   Previous reports have noted potential spin-selective matrix element effects in the vicinity of the Dirac bands that can alter band intensities~\cite{Lin2020, Kang2020b} but the differences in behavior of the Dirac bands above and below the Dirac point is unusual.  

In addition to the intensity suppression of the bands above the Dirac point, the spectral widths of the EDC features above and below the Dirac point decrease as the Dirac point is approached as shown in Fig. S2 in the Supplemental Material~\cite{SM}.  As shown in Fig.~\ref{Theory}, theory predicts several bands that converge at this Dirac point and can affect both the intensity distribution and spectral widths of the observed band structure.  Finally, the background intensity away from the dispersive features is lower above the Dirac point when compared to below the Dirac point.  

Fitting the entire spectra with two Lorentz functions plus a linear background fails to achieve reasonable results.  However, when the $FWHM$ for both Lorentz functions is constrained to the intermediate fitted values from the blue curve in Fig.~\ref{Simulations}d (see Fig. S2~\cite{SM}) then a gap of $2\Delta = 43 \pm 34$ meV results as shown in Fig.~\ref{Simulations}g.  It should be noted that the fitted dispersion deviates from a linear behavior which could be due to the multiple bands with different dispersions in the region.  While the analysis suggests a gapped structure, the empirical observations of variations of the intensity and width of the spectral features in the data raises questions regarding the validity of the gapped structure observed in Fig.~\ref{Simulations}g and ~\ref{Simulations}j.

To investigate how the observed variations in the spectral intensity, width and background can affect EDC fits and the curvature analysis, a phenomenological model is developed to simulate ARPES data where similar analysis can be performed.  Two overlapping bands with linear dispersion are assumed and modeled according to a simple interaction matrix~\cite{Brouet2008,MooreR2010}\[
\begin{vmatrix}
\epsilon_1(k) & \Delta \\
\Delta & \epsilon_2(k)
\end{vmatrix}
\]where $\epsilon_1(k)$ and $\epsilon_2(k)$ are the band energies and $\Delta$ is a momentum independent interaction term. The spectral weight is computed and convoluted with a Lorentz function.  To mimic the reduction of the dispersion widths observed in the data, the $FWHM$ of the Lorentz function is reduced from $0.4$ eV to $0.24$ eV using a parabolic function with the minimum occurring at the Dirac point.  In addition, a similar parabolic function is used to reduce the intensity of the bands above the Dirac point to $70\%$ of their original value at the Dirac point.  These values are chosen based on trends observed in the ARPES data.  Poisson noise is added to the resulting dispersing bands which are then convoluted with a Gaussian with $FWHM = 16$ meV to match the energy resolution used for the ARPES measurements.  Binned results for model data using $\Delta = 40$ meV and $0$ meV are plotted in Fig.~\ref{Simulations}b and ~\ref{Simulations}c respectively.    It should be emphasized that the $\Delta = 40$ meV is manually implemented which forces a gap in the model spectra and is chosen based on theoretical predictions.  Simulating gapped and ungapped spectra with the atypical intenisty and width behavior of the bands near the Dirac point is designed to add credence to the gap analysis of the actual ARPES data.

The model data qualitatively follows the observed trends of the ARPES data and is used to help determine the reliability of the analysis performed on the data.  Fitting the model data with $\Delta = 40$ meV is shown in Fig.~\ref{Simulations}h.  If all parameters are free (blue markers in Fig.~\ref{Simulations}h), then a fitted gap of $2\Delta = 48 \pm 54$ meV results.  If the widths of the two Lorentz functions are constrained during the fit, as was used for the actual data, then a gap of $2\Delta = 46 \pm 38$ meV results.  The fitted gap is smaller than the $2\Delta = 80$ meV used to generate the simulated data.  For simulations using $\Delta = 0$ meV, the fits must be constrained to prevent erratic results and confirms there is no gap in the structure.  

The curvature analysis for the simulated data yields a $2\Delta = 46$ meV gap for the gapped model as shown in Fig.~\ref{Simulations}k, which again is smaller than the gap used to simulate the data.  The reduction of both the intensity and width of the model functions in the simulations yields a suppression of the signal intensity in the curvature plot in agreement with the analysis on the ARPES data but also leads to a reduction of the observed gap.  The curvature analysis of the simulated data with no gap confirms that no gap is observed in the analysis.  These results add confidence to the gapped results in the ARPES data analysis.  The model shows the reduction of intensity and widths observed in the ARPES data tends to reduce the measured gap and thus the gap in the ARPES data may be larger than what is resolved.  The combination of the analysis on the data as well as the simulations allows us to set a conservative gap estimate of $2\Delta\sim60-80$ meV in the ARPES data due to the axial AFM magnetic structure. 

Previous results show a temperature dependent phase transition out of the axial AFM ground state at $T \sim 150$ K~\cite{Meier2019}.  However, as shown in Supplemental Material Fig. S1~\cite{SM}, thermal broadening prevents an accurate determination of the gap closing at elevated temperatures.

\begin{figure*}[t]
\begin{center}
\includegraphics[keepaspectratio=true, width = 7.0 in]{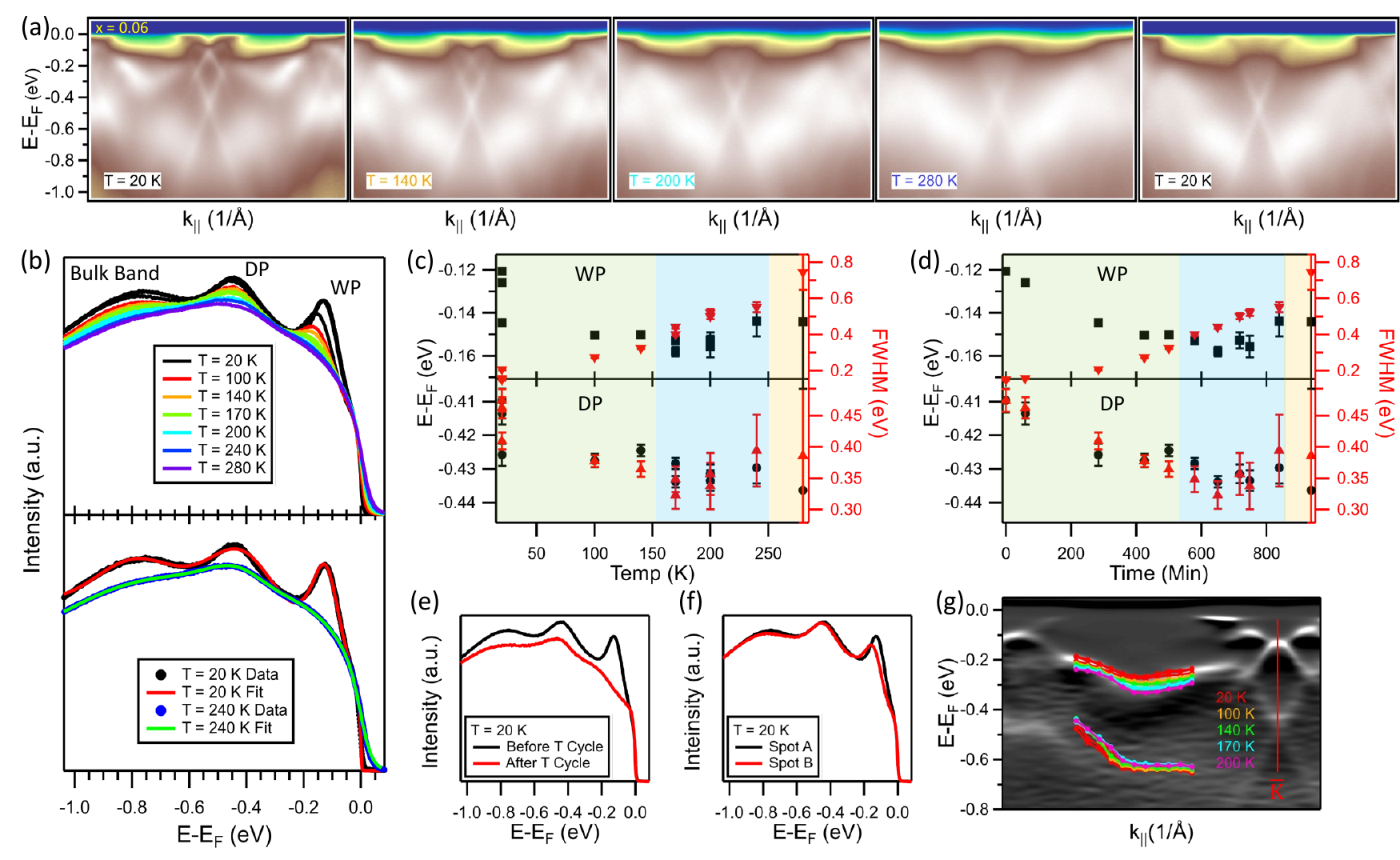}
\caption{Temperature dependent ARPES data and EDC fits for $x = 0.06$.  (a) ARPES data for $x = 0.06$ taken at different temperatures with the rightmost panel taken after cooling back down to $T = 20$ K.  (b) EDCs at the $\bar{\rm K}$ point showing the surface WP, bulk DP and bulk bands for the different temperatures.  Model fits of the data are shown for $T = 20$ K and $240$ K.  (c) Model fit results versus temperature for the position and width of the surface WP and bulk DP.  (d) Model fit results versus data collection time for the position and width of the surface WP and bulk DP.  The planar (green), tilted (blue) and axial (orange) phases are shaded in (c) and (d).  (e) Comparison of EDCs at $T = 20$ K before the temperature cycle and after the temperature cycle highlighting the degradation of the surface WP spectral intensity. (f) Comparison of EDCs at $T = 20$ K before the temperature cycle for two different probed locations on the sample highlighting the sample surface inhomogeneities.  (g) Curvature plot of $x = 0.06$ ARPES data with temperature dependent model fit results overlay for surface and bulk dispersive features away from the $\bar{\rm K}$ high symmetry point. }
\label{TempDep}
\end{center}
\end{figure*}

\subsection{Temperature Dependence for $x = 0.06$}
For $x = 0.06$ the ground state is in the planar magnetic phase.  Magnetization and neutron measurements show a transition to the tilted phase at $T_2 = 256$ K and to the axial phase at $T_1 = 155$ K~\cite{Meier2019}.  To investigate the affects of the changing magnetic moments on the electronic structure, temperature dependent data were taken as shown in Fig.~\ref{TempDep}.  From the plots of the raw data in Fig.~\ref{TempDep}a, general trends can be observed for the topological band dispersions.  Most notably is the downward shift in energy of the surface Weyl point as well as the disappearance of the surface Weyl bands.  The intensity of the bulk Dirac bands decreases with increasing temperature but recovers as temperature is lowered again to $T = 20$ K while the surface Weyl band intensity does not recover when subsequently cooled.  To track the behavior of the bulk Dirac and surface Weyl points the EDC data through the points were fitted with a simple phenomenological model involving Lorentz functions for the bulk Dirac and surface Weyl points as well as a Gaussian function for the bulk bands at $E-E_F \approx -0.8$ eV.  A Shirley background and a Fermi cutoff is included to model the entire EDC as shown in Fig. 3b~\cite{Shirley1972,Castle2001}.    Fit results for different temperatures are shown in Fig.~\ref{TempDep}c.  The fit results show significant scatter in the data with a relatively constant energy for the surface WP and increasing width.  The bulk DP shows a relatively constant energy and width as temperature is increased.  While the increasing width of the surface WP is indicative of a gap opening, the disappearance of the surface Weyl bands decreases the reliability of these fitted widths.  The bulk DP data again shows large scatter with little evidence of a gap structure forming with the changing magnetic moment.  

We found that a more revealing way to approach the analysis is to plot the trends versus time of the data collection as shown in Fig.~\ref{TempDep}d.  Here clear trends become more evident and reveal a monotonic decrease of the WP energy from $E-E_F \sim -0.12$ eV to $E-E_F \sim -0.16$ eV with time along with a factor of four increase in the fit width of the feature.  These trends, along with the disappearance of the surface Weyl bands are indicative of sample aging.  The bulk Dirac bands show a slight decrease in energy with temperature from $E-E_F \sim -0.41$ eV to $E-E_F \sim -0.43$ eV with a large reduction in fit width.  The reduction in width of the feature is opposite the trend expected for an opening gap.   As shown in the bottom panel of Fig.~\ref{TempDep}b, the width and position of the bulk DP feature are slightly skewed at low temperature due to using a simplistic model but the fits of the bulk features improve as the surface feature disappears.  Hence the trends in both the WP and DP width are attributed to the disappearance of the surface Weyl bands.  The conclusion is that surface aging results in the disappearance of the surface Weyl bands.  While no clear evidence of a gap is observed for the bulk Dirac bands, the combination of thermal broadening and shifting of the Weyl bands could obscure the gap opening.

\begin{figure*}[t]
\begin{center}
\includegraphics[keepaspectratio=true, width = 7.0 in]{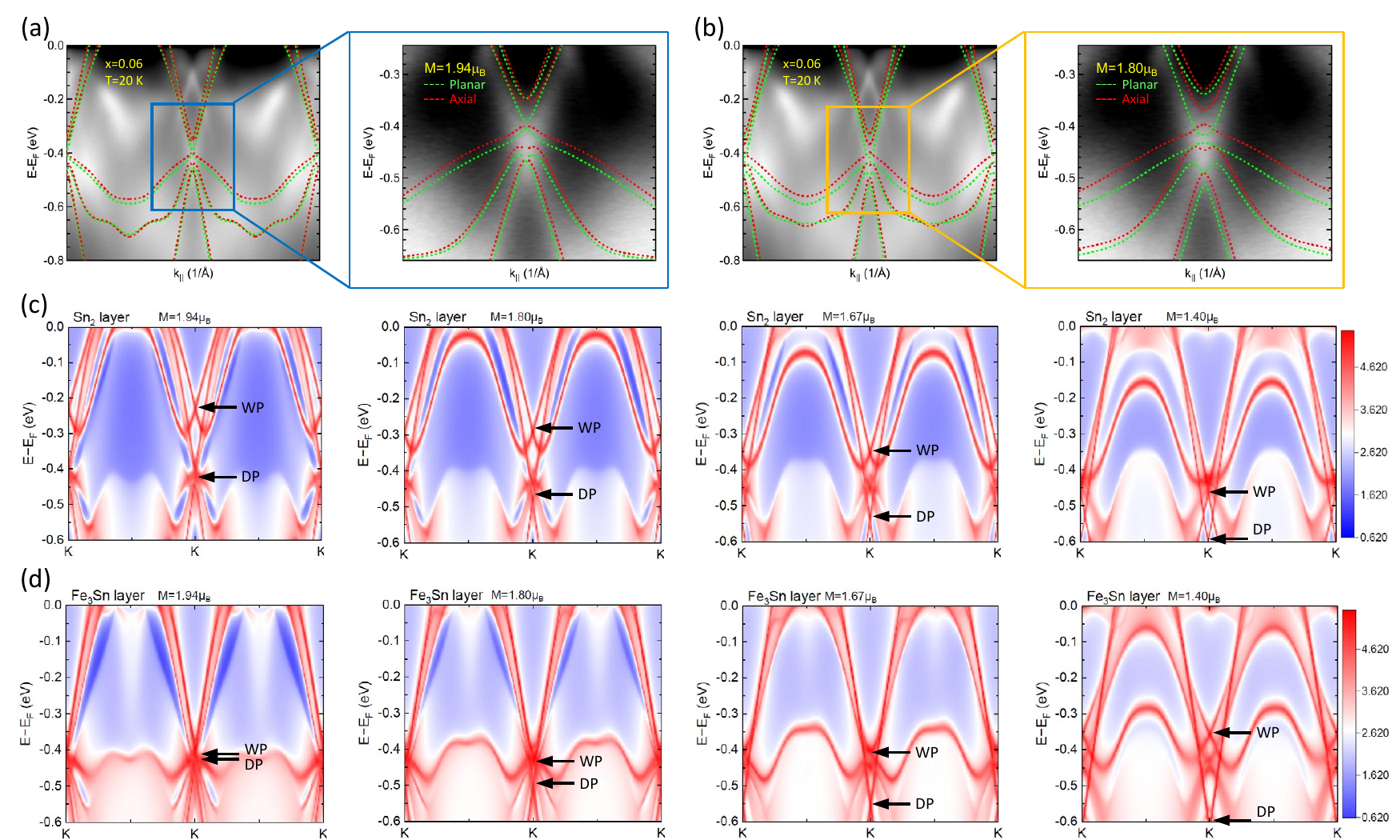}
\caption{Comparison of ARPES with DFT Calculations.   (a) DFT bulk bands with $M = 1.94$ $\rm{\mu_B}$ for planar (green) and axial (red) AFM phases overlaid on ARPES data for $x = 0.06$.  In the zoomed region outlined in blue, the contrast for the ARPES data is adjusted to highlight the Dirac dispersion.  (b) DFT bulk bands with $M = 1.80$ $\rm{\mu_B}$ for planar (green) and axial (red) AFM phases overlaid on ARPES data for $x = 0.06$.  In the zoomed region outlined in orange, the contrast for the ARPES data is adjusted to highlight the Dirac dispersion. (c) DFT slab calculations with Sn$_2$ termination for different $M$.  (d) DFT slab calculations with Fe$_3$Sn termination for different $M$. For (c) and (d) the surface WP and the bulk DP are highlighted for clarity.}
\label{Theory}
\end{center}
\end{figure*}

While the data does not support the opening of a gap due to the reorientation of the magnetic moments, additional trends in the data are observed that give us insight into the band structure evolution with temperature and/or time.  In the low temperature data shown in Fig.~\ref{TempDep}a and its curvature plot in~\ref{TempDep}g, additional bands are observed around $E-E_F = -0.2$ to $-0.3$ eV between the $\bar{\rm K}$ points of neighboring Brillouin zones.  Like the surface Weyl bands, these spectral features appear to move down in energy and disappear as the sample temperature increases.  To better track the trends of these bands, EDCs at $k_{\vert\vert}$ points away from the topological bands were fit with two Lorentz functions to mimic these shallow bands as well as deeper bands around $E-E_F \sim 0.6$ eV.  The fit results are overlaid on the curvature plot of the $T = 20$ K data in Fig.~\ref{TempDep}g.  As the sample temperature increases the shallow bands move down in energy and follow the behavior of the surface Weyl bands while the deeper bands do not move in energy with temperature.  Due to the similarity in behavior with the surface Weyl bands, the shallow bands are attributed to surface states.

It should also be noted that surface inhomogeneities are observed in the ARPES data for different locations probed by the incoming light.  The differences in the EDCs for the $\bar{\rm K}$ points at two different sample locations are shown in Fig.~\ref{TempDep}f.  There are subtle differences in the bulk DP and deeper bulk bands but differences in the energy and intensity of the surface WP are more prominent at the different locations.  In general, inhomogeneities in the surface Weyl bands are observed across the cleaved sample surface while the bulk bands remain consistent.  Due to the $\sim0.04$ mm$^2$ spot size, a single sample spot was isolated for temperature dependent data to ensure variations observed in the bands are due to temperature and not due to surface inhomogeneities.  

\subsection{Comparison with Theory}

To gain a deeper understanding of the observed band structure evolution, DFT calculations including spin-orbit coupling are compared with the ARPES data as shown in Fig.~\ref{Theory} as well as in the Supplemental Material~\cite{SM}.  While the photon energy, $h\nu = 92$ eV, maximizes the intensity within the topological bands, it lies between the H and K points at $k_z \sim 0.48 \pi/c$.  Nonetheless, theoretical calculations reveal a gap opening at this $k_z$ as the spins reorient along into the axial phase, in agreement with previous reports~\cite{Lin2020}.  To understand the implications of spin reorientation and the size of the magnetic moment in the planar and axial phases, constrained magnetic calculations were performed as shown in Fig.~\ref{Theory}a and in the Supplemental Material~\cite{SM}.  Previous reports indicate magnetic moments in the first and bulk Fe$_3$Sn layers to be $2.16$ $\rm{\mu_B}$ and $1.96$ $\rm{\mu_B}$ respectively~\cite{Lin2020}.   There is a large shift of the bulk bands for different magnetic moments as shown in Fig. S3 in the Supplemental Material~\cite{SM} but the theory best matches the ARPES data for $M \geq 1.8$ $\rm{\mu_B}$.  For the larger magnetic moments, the bulk axial and planar band dispersions are similar with the largest difference occurring at the Dirac point as shown in Fig.~\ref{Theory}a and b which agrees with the low temperature ARPES data for $x = 0.06$ in the planar phase.  No shifting or renormalization of the theoretical bands is necessary to match the ARPES data indicative of a lack of strong electronic correlations in the system.  This is to be expected for an itinerant magnetic metallic system and confirms the DFT approach without a Coulomb interaction potential ($+U$) accurately represents the system.

The bulk band calculations do not reveal the Weyl-like bands at $E-E_F \sim -0.1$ eV at the $\bar{\rm K}$ point or bands at $E-E_F \sim -0.2$ eV away from $\bar{\rm K}$ point, implying that these are surface bands induced by the surface Stark effect~\cite{Lin2020}.  To better understand the surface induced bands and their relation to the observed ARPES data, spectral weight from theoretical slab calculations were determined for different magnetic moments in both the planar and axial phases as shown in Fig.~\ref{Theory}c and d and in the Supplemental Material~\cite{SM}.  There are two possible surface terminations, Sn$_2$ and Fe$_3$Sn, and slab calculations are presented for both.  For the larger $M \geq 1.8$ $\rm{\mu_B}$ as previously reported~\cite{Lin2020} only the Sn$_2$ terminated surface yields an observable surface WP shifted up in energy $\sim 0.2$ eV from the bulk DP.  The Sn$_2$ slab calculations for $M = 1.94$ $\rm{\mu_B}$ shows the surface WP at $E-E_F = 0.24$ eV and is similar to the observed WP at $E-E_F = 0.12$ eV in the low temperature $x = 0.06$ ARPES data.  Interestingly, the slab calculations for the Sn$_2$ and Fe$_3$Sn layers show distinct trends in the position of the surface bands as the magnetic moment is varied.  For the Sn$_2$ surface, the surface WP moves down in energy $\sim 0.2$ eV as the magnetic moment is reduced from $M = 1.94$ $\rm{\mu_B}$ to $M = 1.4$ $\rm{\mu_B}$ as shown in Fig.~\ref{Theory}c.  This downward shift in energy with reduced magnetic moment continues for lower $M$ as shown in the Supplemental Material~\cite{SM}.  In contrast, the Fe$_3$Sn surface shows the opposite trends where the surface WP is nearly degenerate in energy with the bulk DP for $M = 1.94$ $\rm{\mu_B}$ and moves upwards in energy $\sim 0.1$ eV as the magnetic moment is reduced to $M = 1.4$ $\rm{\mu_B}$ as shown in ~\ref{Theory}d.  Similarly, this trend continues for lower magnetic moments as shown in the Supplement Material~\cite{SM}.  Since the Fe $d$-states are more concentrated near the Fermi level compared to the extended Sn $p$-states in energy~\cite{Lin2020}, it is speculated that a possible anti-crossing like effect could create the dichotomy in behavior for the two surfaces. However, more efforts are required to confirm such behavior.  Both surface terminations show similar trends in the low energy spectral weight in the region between the $\bar{\rm K}$ points with a downward shift from the Fermi level to $E-E_F \sim 0.1 - 0.15$ eV as the magnetic moment is reduced from $M = 1.94$ $\rm{\mu_B}$ to $M = 1.4$ $\rm{\mu_B}$ as shown in Fig.~\ref{Theory}c and d.

\section{DISCUSSION}

The topological band structure for the bulk Dirac and surface Weyl bands are clearly visible in the ARPES data and are similar to previously published results~\cite{Lin2020}.  Theoretical investigations have predicted that a large $2\Delta \sim70$ meV gap should open at the bulk Dirac point with a reorientation of the magnetic moment out of the kagome lattice plane due to the breaking of the nonsymmorphic $S_{2z}$ symmetry~\cite{Lin2020}.  Previous transport and neutron investigations show the magnetic moment in the (Fe$_{1-x}$Co$_x$)Sn material family can be tuned with composition and temperature~\cite{Meier2019}.  A gap is observed in the low temperature axial phase for $x = 0.17$ but no clear changes to the bulk Dirac bands are observed as the magnetic moments reorient from the planar through the tilted to the axial phase in $x = 0.06$ with increasing temperature.  The widths of the spectral features plus finite energy resolution and thermal broadening of the ARPES measurements may prevent the observation of gap features in the temperature dependent data.  Nonetheless, a large $\sim70$ meV gap is well within the energy resolution of the current measurements and no signatures of such a gap formation are observed for $x = 0.06$ as the system temperature increases through the $T_1$ and $T_2$ magnetic phase boundaries.

Differences are observed in the behavior of the surface Weyl bands compared to the bulk Dirac bands as the sample temperature increase.  The disappearance of the surface bands with temperature is indicative of surface aging, but the downward energy shift of the surface bands only is unusual.  Changes in carrier doping due to aging can alter observed band energies but it is expected that it would alter the energy of all the observed bands and not just the surface bands.  

Theoretical calculations not only confirm a large gap should appear due to the breaking of $S_{2z}$ symmetry but also reveal a large energy shift of the bulk bands away from the K-point as the c-axis magnetic moment changes, which is not observed in the ARPES data.  Contrary to bulk band behavior in the ARPES data, the surface bands clearly shift downwards in energy as the temperature and/or time of measurement increases.  Theoretical slab calculations for the Sn$_2$ termination agree with the ARPES data in the energy of the surface WP and also shows a trend downward in energy as the magnetic moment is reduced.  

In addition to the temperature trends in the surface bands, variations in the surface WP are observed due to surface inhomogeneities across the sample surface.  For FeSn the magnetic anisotropy energy is shown to be small, on the order of $~0.03$ meV/unit cell, where small perturbations can manipulate the spin orientation~\cite{Lin2020,Smejkal2017,Smejkal2018}.  This anisotropy contributes to the tunability of the magnetic moments but can also contribute to a large magnetic inhomogeneity due to different terminations and defects on the surface.   The spatial variations of the surface Weyl bands are likely due to inhomogenietes of the magnetic moments on the surface resulting in variations of the observed band energies.  

Similarly, shifts in energy as the sample surface ages is likely due to a reduction or disordering of the magnetic moments due to surface aging effects.  The differences in the surface potential which result in the appearance of the surface Weyl bands also results in alterations of the magnetic moments of the surface most Fe$_3$Sn layer being isolated to the surface Weyl bands and not affecting the bulk Dirac bands.  Hence the observed behavior of the surface Weyl bands with temperature/time and variations in sample position suggest the magnetic moments in the surface most Fe$_3$Sn layer reorient or relax during the ARPES measurements.  Surface magnetic measurements are necessary to confirm the magnetic moments on the surface as well as their disorder/relaxation over time.

In conclusion, a sizable $2\Delta \sim 60-80$ meV gap at the bulk Dirac point is observed in the low temperature axial phase for $x = 0.17$ but no clear gap opening is observed with temperature across the planar to axial phase transition in the $x = 0.06$ samples.  However, large energy shifts of the surface bands are observed.  The shift, and eventual disappearance, of the surface bands are attributed to a reduction and disordering of the surface magnetic moment as the sample ages after cleaving, which is an unavoidable artifact of the surface sensitive experimental technique.   The combination of thermal broadening and shifting of the surface Weyl bands could obscure the gap evolution with temperature and more systematic studies at different doping levels are required to reveal the subtle details.  The conflicting gap results for the two dopings are on par with investigations of other similar magnetic topological systems and highlights the challenges in disentangling magnetic and topological features in these material systems.  Nonetheless, the shifting surface Weyl bands reveal the need to carefully track and account for subtle changes in surface magnetic moments during measurements. Such accounting is critical to understand their influence on surface sensitive probes and could account for discrepancies observed in published results for related systems.  Future efforts are necessary to understand the evolution of the surface magnetism but clear links between the topological electronic structure and magnetic moment are evident as our work has revealed.

\begin{acknowledgements}
The work of all coauthors was supported by the U.S. Department of Energy, Office of Science, Basic Energy Sciences, Materials Sciences and Engineering Division.  The use of the Stanford Synchrotron Radiation Lightsource, SLAC National Accelerator Laboratory, is supported by the U.S. Department of Energy, Office of Science, Office of Basic Energy Sciences, under Contract No. DE-AC02-76SF00515.  This research used resources of the Compute and Data Environment for Science (CADES) at the Oak Ridge National Laboratory, which is supported by the Office of Science of the U.S. Department of Energy under Contract No. DE-AC05-00OR22725.
\end{acknowledgements}

%


\end{document}